\documentclass[sigconf]{acmart}

\AtBeginDocument{%
  \providecommand\BibTeX{{%
    \normalfont B\kern-0.5em{\scshape i\kern-0.25em b}\kern-0.8em\TeX}}}

\acmConference[SBES'24]{Brazilian Symposium on Software Engineering}{September 30 -- October 04, 2024}{Curitiba, PR}

\usepackage{enumitem}
\setlist[itemize]{noitemsep, topsep=0pt}

\usepackage[normalem]{ulem}
\usepackage{overpic,color}
\usepackage{pgf-pie}  
\usepackage{tikz}
\usepackage{pgfplots}

\pgfplotsset{compat=1.8}
\usepgfplotslibrary{statistics}
\usepackage{pgfplotstable}

\usepackage{wrapfig}

\usepackage{algorithmic}
\usepackage{graphicx}
\usepackage{textcomp}
\usepackage{adjustbox}
\usepackage{booktabs}
\usepackage{float}
\usepackage{xcolor}
\usepackage[skins]{tcolorbox}

\usepackage{hyperref}
\hypersetup{
    colorlinks=true,
    linkcolor=blue,
    anchorcolor=blue,
    filecolor=blue,      
    citecolor=blue,
    urlcolor=blue
}

\usepackage{balance}
\usepackage{tabularx}
\usepackage{graphicx}
\usepackage[gen]{eurosym}

\usepackage{subfig}

\usepackage{lipsum} 

\usepackage{soul}
\setstcolor{red}

\newcommand{\tname}{\textsc{R-DLD}} 

\newcommand{\myparagraphtwo}[1]{\vspace{0.5ex}\noindent \textbf{#1 }}
\newcommand{\myparagraphitalic}[1]{\vspace{0.5ex}\noindent \textit{#1 }}

\newcommand{\Ignore}[1]{}

\newcommand{\etal}{{\em et al.\/}}

\newcommand{\upd}[1]{{#1}}

\newcommand{\totalNumAppsCandidate}{\upd{733}}
\newcommand{\totalNumAppsAnalyzed}{\upd{77}}
\newcommand{\totalNumAppsWithBug}{\upd{67}}

\newcommand{\totalhoras}{\upd{350}}

\newcommand{\totalNumWarnings}{\upd{3,589}}

\newcommand{\totalNumTrueBugs}{\upd{2,160}}
\newcommand{\totalNumFalseAlarm}{\upd{1,429}}
\newcommand{\totalBugsWithoutDuplication}{\upd{341}}
\newcommand{\totalBugsWithoutDuplicationNotArchived}{\upd{336}}

\newcommand{\percentualDeTrueBugsTotal}{\upd{60.2\%}}
\newcommand{\percentualDeFalseAlarmTotal}{\upd{39.8\%}}

\newcommand{\perctPropertyTotalTrueBug}{\upd{18.4\%}}
\newcommand{\perctSnapTotalTrueBug}{\upd{9.7\%}}
\newcommand{\perctBothTotalTrueBug}{\upd{71.9\%}}

\newcommand{\perctPropertyTotalFalseAlarm}{\upd{2.0\%}}
\newcommand{\perctSnapTotalFalseAlarm}{\upd{52.8\%}}
\newcommand{\perctBothTotalFalseAlarm}{\upd{45.2\%}}

\newcommand{\perctPropertyTotalFalseAlarmRotation}{\upd{1.7\%}}
\newcommand{\perctSnapTotalFalseAlarmRotation}{\upd{44.9\%}}
\newcommand{\perctBothTotalFalseAlarmRotation}{\upd{44.1\%}}
\newcommand{\perctTotalFalseAlarmRotationIncompleta}{\upd{9.3\%}}

\newcommand{\totalCincoAppAnimationFalseAlarm}{\upd{1,169}}
\newcommand{\totalCincoAppAnimationTrueBugs}{\upd{72}}

\newcommand{\proporcaoTrueBugGeral}{\upd{1.38}}
\newcommand{\proporcaoTrueBugWithoutAnima}{\upd{6}}

\newcommand{\coberturaPriQuartil}{\upd{40\%}}
\newcommand{\coberturaMediana}{\upd{70\%}}
\newcommand{\coberturaTrueBug}{\upd{351}}
\newcommand{\coberturaFalseAlarm}{\upd{470}}
\newcommand{\coberturaFalseAlarmOutlier}{\upd{379}}
\newcommand{\coberturaTrueBugOutlier}{\upd{1}}
\newcommand{\coberturaAppOutlier}{\upd{Babydots}}

\newcommand{\issueTotal}{\upd{87}} 
\newcommand{\issueAnswered}{\upd{52}}

\newcommand{\issuefalhasConfirmadas}{\upd{180}}

\newcommand{\issueAnsweredPercentage}{\upd{58.94\%}}

\newcommand{\issueAnsweredAcceptedPercentage}{\upd{89.55\%}}

\newcommand{\issueAccepted}{\upd{47}}
\newcommand{\issueAcceptedFix}{\upd{18}}

\newcommand{\issueAcceptedWontfix}{\upd{4}}

\newcommand{\issueAcceptedOpen}{\upd{25}}

\newcommand{\issueRejected}{\upd{5}}

\newcommand{\percentagefalhasCorrigidas}{\upd{35.82\%}}

\definecolor{mygreen}{rgb}{0,0.6,0}
\definecolor{mygray}{rgb}{0.5,0.5,0.5}
\definecolor{mymauve}{rgb}{0.58,0,0.82}

\definecolor{lightgray}{gray}{0.9}

\newcommand{\artifactUrl}{\url{https://github.com/RoboticsRG/R-DLD}}

\newcommand{\CodeIn}[1]{\texttt{#1}}

\setcopyright{none} 
\renewcommand\footnotetextcopyrightpermission[1]{} 

\begin{document}
\title{Robotic-supported Data Loss Detection in Android Applications}

\author{Davi Freitas}
\orcid{0009-0000-1419-5206}
\affiliation{%
  \institution{Federal University of Pernambuco}
  \city{Recife}
  \country{Brazil}
}
\email{dsf3@cin.ufpe.br}

\author{Breno Miranda}
\orcid{0000-0001-9608-9393}
\affiliation{%
  \institution{Federal University of Pernambuco}
  \city{Recife}
  \country{Brazil}
}
\email{bafm@cin.ufpe.br}

\author{Juliano Iyoda}
\orcid{0000-0001-7137-8287}
\affiliation{%
  \institution{Federal University of Pernambuco}
  \city{Recife}
  \country{Brazil}
}
\email{jmi@cin.ufpe.br}

\begin{abstract}
Smartphones have become integral to modern life due to their diverse applications. 
However, the development of applications for these devices faces significant challenges, primarily due to two factors: (i) the diverse hardware and operating system versions that require testing across multiple configurations; and (ii) the imperative activity of testing the smartphone in a non-invasive way (similar to an end-user interaction), which is expensive, tedious and error-prone as it is usually carried out manually.
In this context, robotic automation has emerged as an effective solution for addressing the diversity of smartphone systems and versions, and in performing some testing tasks non-invasively. In addition, automated robots facilitate quick, accurate, and repetitive testing, thus enabling developers to validate their applications across various configurations effectively. This results in a less invasive verification and in a significant reduction in time spent on manual testing, thus accelerating the development cycle.
Our work proposes R-DLD (Robotic-supported Data Loss Detection), a robot-assisted infrastructure for data loss detection in Android applications, offering less invasive and more realistic tests by interacting directly with smartphone sensors. The robot is constructed using cost-effective materials, facilitating its adoption in testing environments. In our empirical evaluation \tname{} successfully identified \totalBugsWithoutDuplication{} data loss issues in \totalNumAppsAnalyzed{} randomly selected apps from an Android store. All reported bugs received responses from the developer, with \issueAnsweredAcceptedPercentage{} confirming the data loss problems, while \percentagefalhasCorrigidas{} have being subsequently fixed.

\end{abstract}
\maketitle

\section{Introduction}
\label{sec:introduction}

Smartphones have become an essential tool in peoples lives due to their portability, easy-to-use, and the availability of applications and features that enhance daily life in various dimensions (work, entertainment, shopping, education and communication). 
However, developing high-quality applications for smartphones presents significant challenges in testing. One of the main challenges is the great diversity of hardware and operating system versions, which requires developers to test their applications in multiple configuration environments~\cite{fragmentation}. 
Another challenge is the imperative need to run test cases in an environment that closely mirrors the production setup. This involves using real physical devices, exploring sensors, and ideally, conducting \textit{non-invasive} testing procedures.

For addressing these challenges, test automation emerges as a practical solution. 
Testing frameworks such as Robotium~\cite{Robotium}, Appium~\cite{Appium}, UIAutomator~\cite{UIAutomator}, or even the direct use of Android Debug Bridge (ADB)~\cite{ADB} are very useful for automating the testing process and enhancing efficiency. However, despite their usefulness, they are invasive approaches that simulate screen touches and gestures by sending events via software, under the hood. While these methods offer valuable insights into the functionality of Android applications, their invasive nature may overlook issues that could arise in a non-simulated, physical interactions with the device and its sensors. Non-invasive testing, on the other hand, if conducted manually, is expensive, tedious, and error-prone.

In this context, one promising solution for addressing the challenges of diversity of systems and invasive testing involves the deployment of robotic arms.
As robots are employed in an extensive range of applications and in numerous repetitive tasks, there is a growing interest in their adoption in the context of software testing~\cite{FaceAuthentication, ImageRectification, RoScript, RobotDedicated, JitterReduction, IsomorphicGUI, MultiTouchArduino}.
In particular, Mao \etal{}~\cite{AxizBlackBoxAutomation} advocate for the use of robotics in mobile device testing, asserting that it offers a form of black-box testing that is ``\textit{more black-box than anything witnessed}'' thus far. Using a robotic arm greatly emulates real user and mobile device interaction without requiring any changes to the source code of the application under test.

Recently, Riganelli and colleagues~\cite{DLD} introduced Data Loss Detector (DLD), a technique for revealing \textit{data loss} defects in Android. 
In Android applications, data loss occurs when information is accidentally deleted or when state variables are inadvertently assigned to default or initial values.
Data loss problems are directly related to the life cycle of the Activity component, which is responsible for implementing the application functionalities through different states.
Depending on the amount of resources and the type of event received, the Activity can be temporarily destroyed in order to free resources, and can be later rebuilt.
Such events, called {\em stop-start\/} events, occur frequently and can result in data loss issues. As an example, a rotation of the mobile phone that changes its orientation from portrait to landscape produces a {\em stop-start\/} event.

Inspired by DLD,
and motivated by the need to test mobile devices in an environment that is as close as possible to the production setup,
this paper introduces a robot-assisted infrastructure for data loss detection in Android applications
called \tname{} (Robotic-supported Data Loss Detection). 
Unlike DLD, which simulates sensor inputs to induce device screen orientation changes, \tname{} takes a less invasive and more realistic approach by exercising the actual mobile phone sensors for the orientation changes.
Our infrastructure was constructed using affordable materials, including an Arduino board, a stepper motor, a power source, and a chassis. This design choice aims to facilitate other testing teams interested in replicating our setup.

In order to evaluate our proposed approach, we conducted a conceptual experimental replication of the original DLD study~\cite{DLD} by adapting the research questions to our new context with the robotic infrastructure.

The key contributions of this work include:
\begin{itemize}
    \item The introduction of a robotics-supported approach for detecting data loss in Android apps, extending the Data Loss Detector (DLD) technique~\cite{DLD}.
    \item An evaluation of the proposed approach structured as a conceptual replication of the original DLD evaluation.
    \item The discovery of \totalBugsWithoutDuplication{} new data loss defects across \totalNumAppsWithBug{} apps, with \issueAnsweredAcceptedPercentage{} of them being confirmed by the developer, and \percentagefalhasCorrigidas{} of them being fixed by the time of this submission.
    \item A comprehensive replication package that includes all artifacts produced during our work, aimed at facilitating independent verification and replication of our results\footnote{\artifactUrl{}}.
\end{itemize}

The paper is structured as follows. 
Section~\ref{sec:background} provides background information on data loss defects in Android apps.
In Section~\ref{sec:robot}, we introduce the proposed robotics-supported approach for data loss detection.
Section~\ref{sec:methodology} outlines the methodology employed in our study, while Section~\ref{sec:analysis} presents the analysis of the results of our empirical evaluation.
Section~\ref{sec:feasibility} discusses the feasibility of robotic automation for data loss testing.
Section~\ref{sec:threats} addresses potential threats to the validity or our study.
Section~\ref{sec:related} explores related work and, finally,
Section~\ref{sec:conclusion} summarizes the findings and concludes the paper.

\section{Background}
\label{sec:background}

\subsection{The Activity Lifecycle in Android Devices}

An activity is an entry point for the interaction of an Android application with the user~\cite{AndroidDevelopers}. Typically, an application has multiple activities: a main one that is initiated when opening the app, and secondary activities designed to perform different actions such as configuring a feature or accessing an external service.

An activity is a component with a well known lifecycle. From the moment an application is launched until its destruction, instances of activities transition through various states and execute specific callbacks. The Activity class provides a set of seven callbacks: \mbox{onCreate()}, onStart(), onResume(), onPause(), onStop(), \mbox{onDestroy()} and onRestart().

Each callback  performs specific tasks appropriate to a particular change in the application state. Implementing the callbacks correctly avoids disrupting the user experience with crashes, data loss, or performance issues. However, it is not necessary to implement all methods in the lifecycle. While implementing onCreate() is mandatory, implementing other methods depends on the complexity and behavior of the application. As the activity enters a new state, the system invokes each of these callbacks.

\myparagraphtwo{onCreate()} is triggered as soon as the system creates the activity and enters the "Created" state. This callback must be implemented and performs the basic initialization of the application. It receives the parameter savedInstanceState, a Bundle object containing the previously saved state of the activity.

\myparagraphtwo{onStart()} is triggered when the activity enters the Started state, making the activity visible to the user.

\myparagraphtwo{onResume()} is triggered when the activity enters the Resumed state, where the application is in the foreground, and the user can interact with it. The activity remains in this state until an interruption occurs (switching between apps, being interrupted by another action, etc.), which puts it in the Paused state. If the app returns to the foreground again, it goes to the Resumed state.

\myparagraphtwo{onPause()} is called when the activity moves to the Paused state, indicating that the user is leaving the activity. It is typically used to release resources not in use, like a camera or a sensor. However, lengthy operations, such as data saving, should be avoided as they may not finish before the method exits.

\myparagraphtwo{onStop()} is triggered when the activity enters the Stopped state. It is a suitable callback to implement data saving or releasing unnecessary resources. From the Stopped state, the activity either interacts with the user again or is terminated. If the activity returns to the foreground, the system invokes onRestart(); otherwise, it calls onDestroy().

\myparagraphtwo{onDestroy()} is triggered before the activity is destroyed. This occurs when the user finishes the activity and it is completely discarded, or when the system temporarily destroys the activity due to a configuration change, such as a layout change due to a device orientation change. During the orientation change, the activity is temporarily destroyed and then restarted to accommodate its elements in the new device orientation.

\myparagraphtwo{onRestart()} is triggered after the onStop() callback when the activity goes out of focus and is being re-displayed to the user. It is followed by the onStart() callback.

The configuration change triggered by the stop-start event caused by an orientation change has a higher probability of identifying a data loss problem, as the activity is destroyed by the onDestroy() callback and then recreated with the new device orientation. If the developer has not implemented methods to save the state of variables or has not correctly implemented lifecycle-aware components, they will be restarted with default values, resulting in data loss.

\subsection{Data Loss Detector (DLD) Tool}
\label{sec:dld}

The Data Loss Detector (DLD)~\cite{DLD} is a tool designed to detect data loss issues in Android applications based on DroidBot~\cite{DroidBot}. It identifies data loss problems caused by changes in the smartphone orientation, thus triggering the stop-start event.

DroidBot~\cite{DroidBot} is a user interface-guided test input generator for Android applications. Since input generation is guided by the user interface, there is no need to instrument the application or have access to the source code.

\begin{figure}[htbp]    
    \centerline{\includegraphics[width=1\columnwidth]{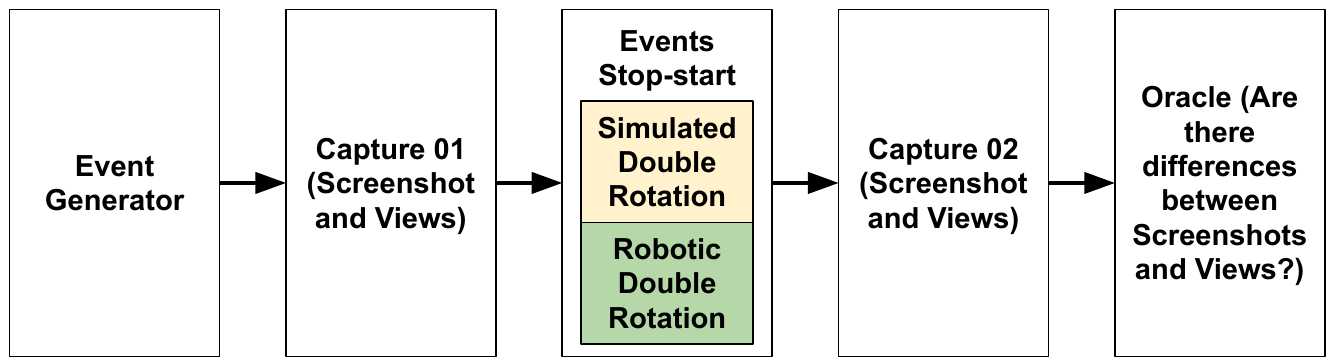}}
    \caption{DLD/RDLD process for detecting data loss\label{fluxo_dld_rdld}}
\end{figure}

DLD utilizes exploratory strategies and actions that maximize the likelihood of finding data loss defects through double rotation, which forces an activity destruction caused by the change in orientation.
Then, the DLD oracle employs two approaches to identify data loss. This process occurs in five steps (Figure~\ref{fluxo_dld_rdld}).

The first step generates a set of events that modifies the default values of the elements in the activity, for example, by changing a field value (via typing) from its default value zero to the value, say, three. Then, a screenshot of the activity is taken and the views (properties of the screen elements) are saved.

\begin{figure}[htbp]
    \subfloat[Portrait]{{\includegraphics[width=0.14\columnwidth]{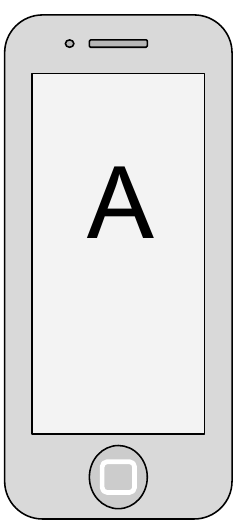} }}%
    \enspace
    \subfloat[Inverted\\portrait]{{\includegraphics[width=0.14\columnwidth]{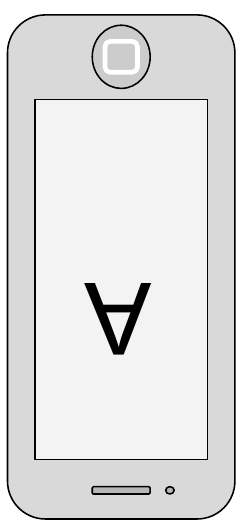} }}%
    \enspace
    \subfloat[Left\\landscape]{{\includegraphics[width=0.25\columnwidth]{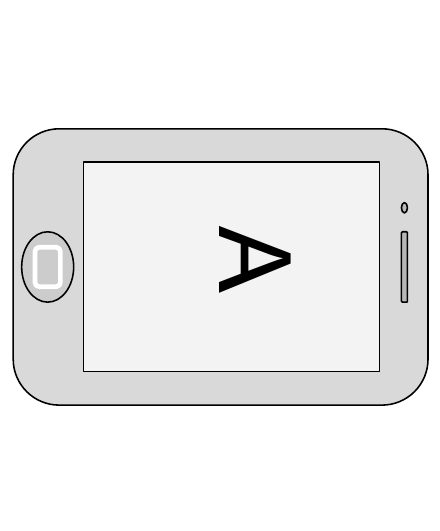} }}%
    \enspace
    \subfloat[Right\\landscape]{{\includegraphics[width=0.25\columnwidth]{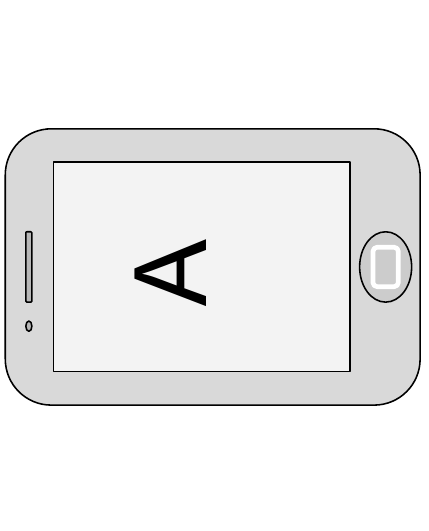} }}%
    \caption{Smartphone orientations. A double rotation goes from (a) to (c), and back to (a).\label{fig_orientation}}  
\end{figure}

Next, a double rotation is performed  using ADB commands that move the smartphone from the portrait position (Figure~\ref{fig_orientation}(a)) to the left landscape orientation (Figure~\ref{fig_orientation}(c)), and then back to portrait (Figure~\ref{fig_orientation}(a)). The double rotation is used to compare screen captures in the same orientation. Following this, a new state screenshot is taken but, this time, after the double rotation stop-start event that has forced the destruction and recreation of the activity. DLD performs this double rotation by simulating it in software (highlighted by the yellow box in Figure~\ref{fluxo_dld_rdld}).

Finally, the oracle checks for data loss by comparing both the screenshots and the views. If there are discrepancies between any of these artifacts, a data loss alert is generated along with the collected information.

The execution of the tool is performed automatically once the user defines several parameters: the target application, the location for the output artifacts, the number of events, and the device under test. The latter can be either an Android Virtual Device (AVD) or a real device and, in both cases, the double rotation is simulated via the Android Debug Bridge (ADB) commands.
The test execution duration depends on the number of events passed as a parameter and the device used (AVD or real device). For instance, the experiment reported in the DLD paper~\cite{DLD} used 2,250 events per app, an AVD and a runtime of 3 hours.

In what follows, we describe the three methods employed by DLD in order to increase the likelihood of identifying a data loss issue.

\myparagraphtwo{A biased model-based exploration strategy.} This strategy constructs a model of the graphical interface to represent visited states and executed actions. The exploration visits new states and incrementally tests newly discovered ones. DLD generates five types of actions during exploration:

\myparagraphitalic{(a) TouchEvent,} which performs a tap on a clickable view.

\myparagraphitalic{(b) LongTouchEvent,} which performs a long tap on a clickable view.

\myparagraphitalic{(c) SetTextEvent,} which inputs text into an editable view.

\myparagraphitalic{(d) KeyEvent,} which presses a navigation button.

\myparagraphitalic{(e) ScrollEvent,} which performs a swipe on a scrollable view.

\myparagraphtwo{Data loss revelation actions.} The actions are grouped into two types: the ones that are executed systematically every time a new state is reached, and those that are executed probabilistically at each state. When a new state is discovered, systematic data loss revelation actions are performed. Otherwise, a probabilistic data loss revelation action is executed, prioritizing events not executed in the activity. The systematic data loss revelation actions consist of five steps:

\myparagraphitalic{(a) Fill-in}: Interacts with all elements of the activity to input non-empty and non-default values.

\myparagraphitalic{(b) Save state}: Saves the current state of the activity to later check for data loss.

\myparagraphitalic{(c) Double screen rotation}: Performs a screen orientation change in order to produce a stop-start event and thus force the recreation of the activity. Two rotations are performed to return to the initial orientation. The screen, after the double rotation, should be exactly the same as the initial screen. Otherwise, data loss is considered to have occurred.

\myparagraphitalic{(d) Check state}: Compares the current state with the saved state to determine if any data loss occurred.

\myparagraphitalic{(e) Scroll down}: Executes a scroll action that may reveal new elements that reach new states.

\myparagraphtwo{Use of two oracles for data loss detection.} DLD employs the strategy of capturing the state before and after executing actions that may identify data loss issues, and then comparing them. It defines two oracle strategies that can be used independently or jointly: 

\myparagraphitalic{(a) Screenshot-based oracle:} DLD takes a screenshot and crops the header and footer of the image to eliminate time-changing information such as time and battery level. Comparison of the images representing states before and after double rotation is performed pixel by pixel. In order to reduce occurrences of false positives related to cursor blinking, for example, differences up to 15 pixels per 10,000 are disregarded.

\myparagraphitalic{(b) Property-based oracle:} DLD retrieves all views, including their properties and hierarchical organization, into a Python dictionary. The comparison is performed between the dictionaries captured before and after the double rotation.

\section{The \tname{} Infrastructure}
\label{sec:robot}
\tname{} is a robotic device built on the Arduino platform \cite{arduino}, aiming to support the four types of smartphone orientations depicted in Figure~\ref{fig_orientation} (Portrait, Inverted Portrait, Left Landscape, and Right Landscape). \tname{} is activated via a serial communication, and different components from the platform were used: a part consisting of pre-fabricated modules ({\em shields\/}) and another part developed by us in order to integrate them. 

\begin{figure}[htbp]    
    \subfloat[\tname{} infrastructure\label{fig_robo}]{{\includegraphics[width=0.46\columnwidth]{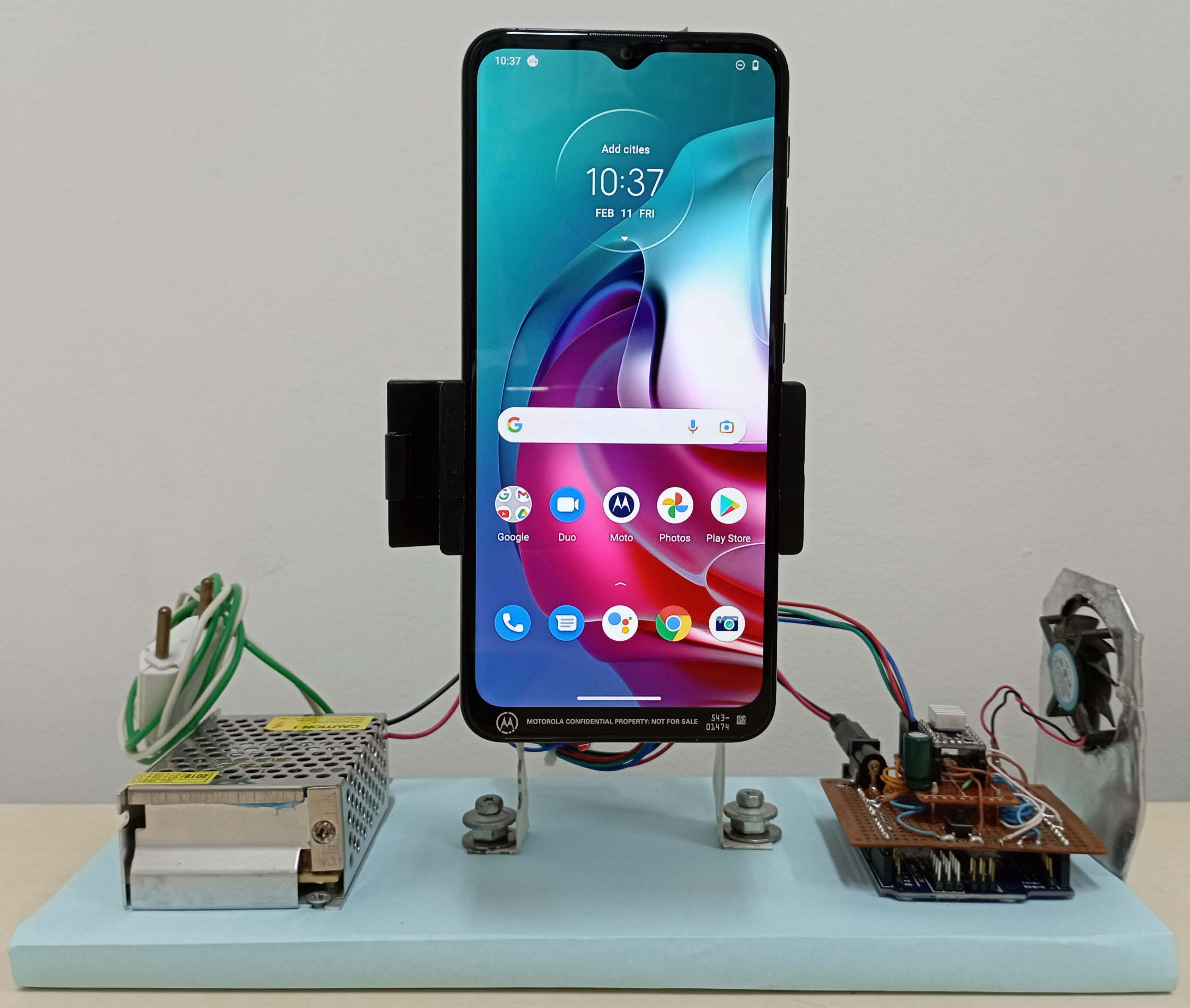} }}%
    \subfloat[Electrical schematic\label{img_esquema}]{{\includegraphics[width=0.54\columnwidth]{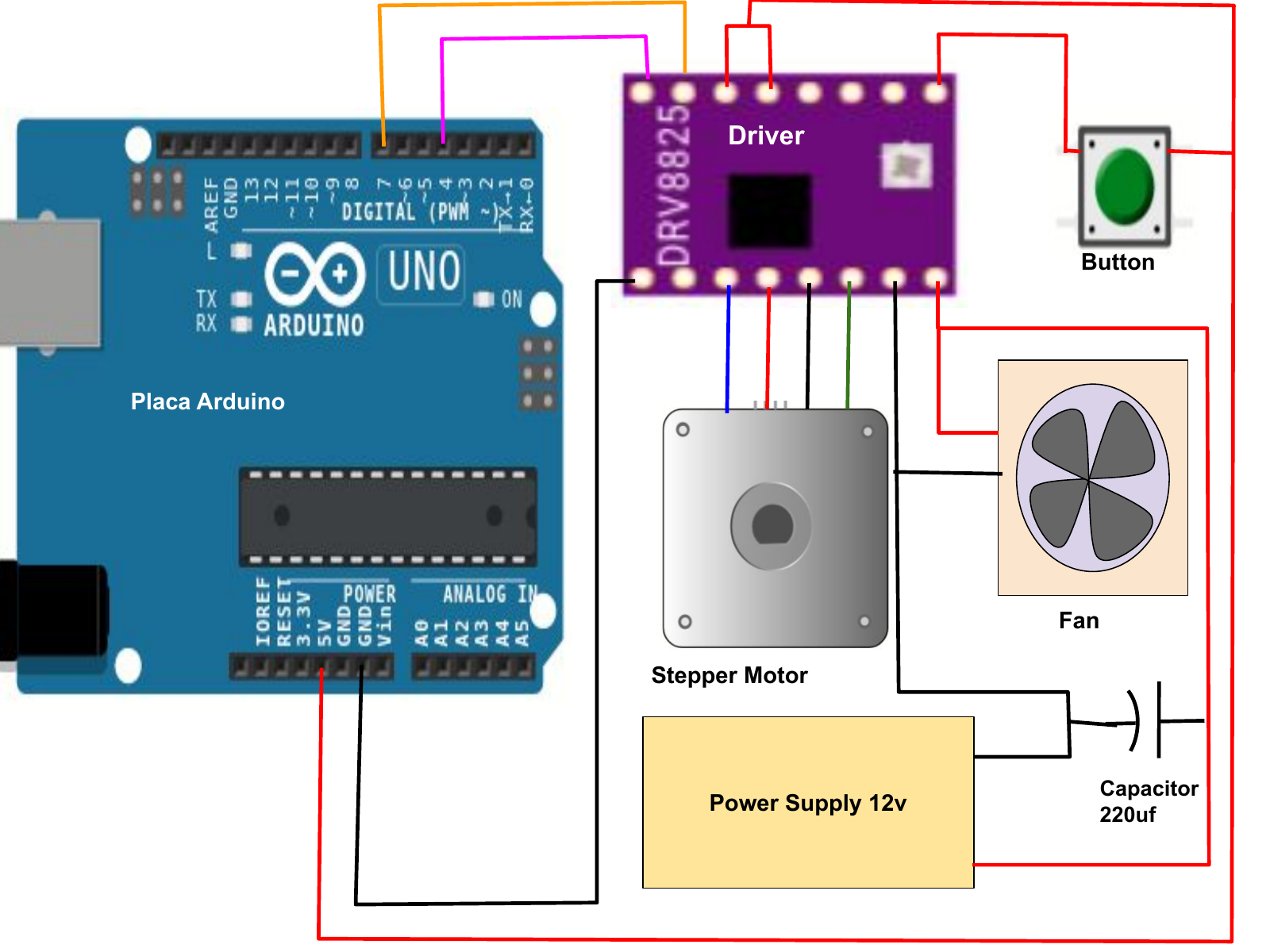} }}%
    \caption{\tname{} infrastructure and Electrical schematic}    
    \label{fig:robo_esquema}
\end{figure}

In order to assemble the robot, the following items were purchased: an Arduino UNO board, a 12-volt 3-watt power supply, a NEMA 17 model stepper motor, and a DRV8825 motor driver. 
Some components had to be fabricated, such as the clamp to hold the smartphone and the robot chassis. The clamp was made using a smartphone car holder, and a shaft was adapted to connect it to the stepper motor. The chassis was crafted in wood and a metal structure to support the components and to ensure stability during rotation.
Figure~\ref{fig:robo_esquema} 
displays the robotic structure (Figure~\ref{fig:robo_esquema}(a)), emphasizing the details of the smartphone-holding grip, 
and depicts the electrical diagram of the component connections (Figure~\ref{fig:robo_esquema}(b)).

A C-based program~\cite{programacao_arduino} was implemented for a computer to communicate with \tname{} via a serial interface in order to position the smartphone in the desired orientation during the data loss test. The Arduino development environment was used along with the AccelStepper library \cite{accelstepper}. Since the stepper motor does not accept angle parameters, the number of steps for each orientation was calculated. The employed motor requires 200 steps for a full revolution, translating to 1.8 degrees per step. Therefore, to achieve the Portrait, Landscape Left, Inverted Portrait, and Landscape Right orientations (Figure~\ref{fig_orientation}), 0, 50, 100, and 150 steps are needed, respectively. The initial position was set as Portrait and can be manually calibrated while pressing the button on the board.

\subsection{\tname{} Architecture \& Modifications to DLD}
\label{sec:architecture}

\begin{figure}[htbp]    
    \centerline{\includegraphics[width=1\columnwidth]{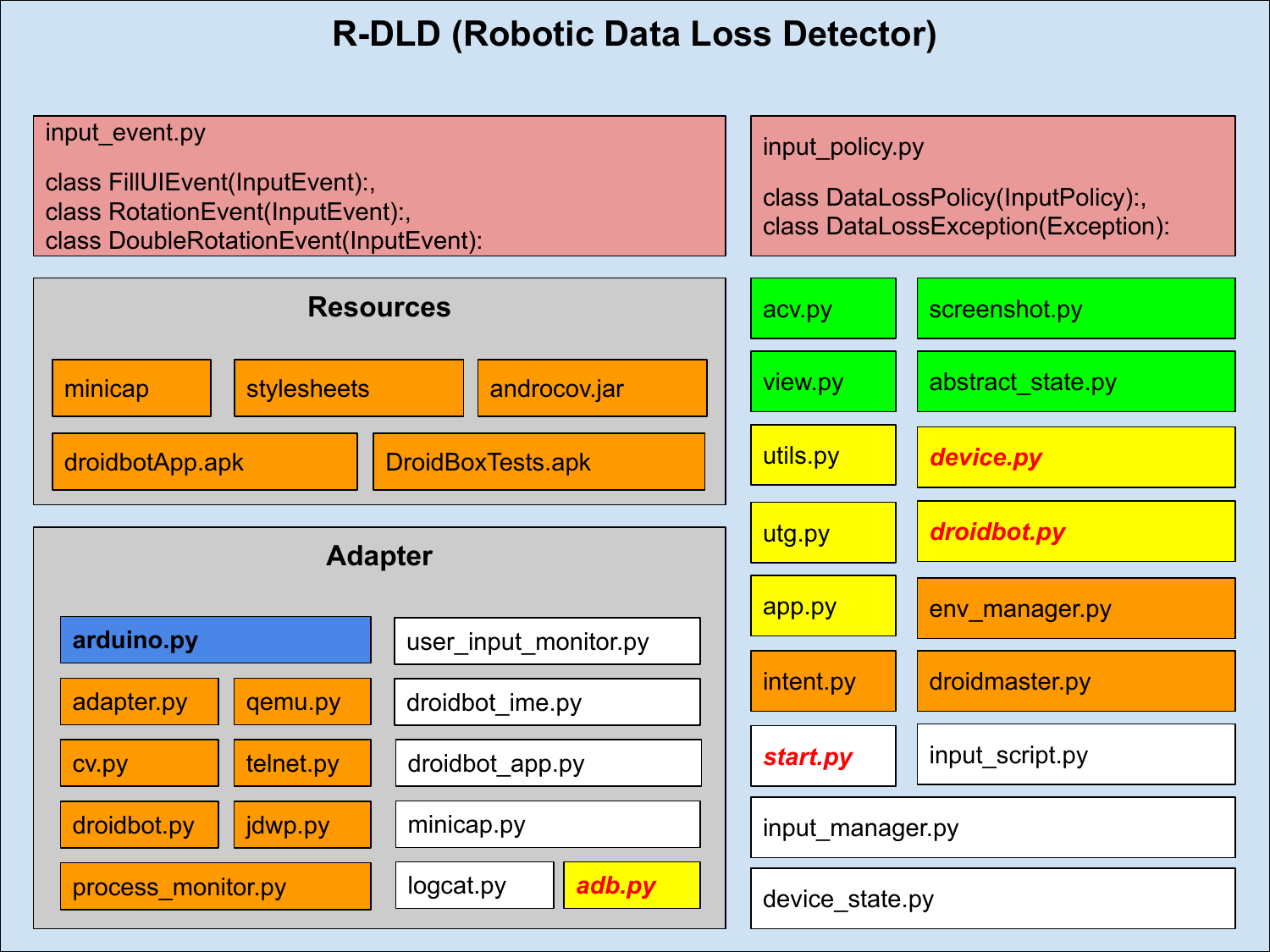}} 
    \caption{Architecture of \tname{} and Modifications to DLD\label{arq_rdld}}
\end{figure}

\tname{} is a robotic extension of DLD incorporating Droidbot and enhancing its data loss detection capabilities. In the architecture depicted in Figure~\ref{arq_rdld}, we can identify the Droidbot original components (in orange), the additions and modifications made by the DLD team (in green, pink, yellow and white), as well as the extensions made to implement \tname{} (in blue).

Considering Droidbot as the core, we explain how DLD and later \tname{} introduced modifications. Droidbot comprises the Adapter module, a Resources folder, and various classes and scripts. The Adapter provides an abstraction for the device and the system under test (SUT). The Resources folder contains auxiliary files, while the classes and scripts control events and inputs during testing.

Modifications added by DLD are highlighted in green, showing the files integrated by the DLD team into Droidbot. Pink files include classes added by DLD to handle inputs and data loss policies. Yellow files introduce methods to aid in data loss detection. The white color indicates minor changes for keeping compatibility with the DLD modifications, while orange files are those that remained unchanged (i.e. the original Droidbot).

\tname{} introduced a new file (in blue) in order to establish a serial communication with the Arduino board. Files named in red have been slightly adjusted by us to incorporate robotic functionalities and the logic to intercept orientation changes sent from DLD to the robot. The \CodeIn{-robot} parameter was created to enable robotic capabilities in \tname{}. When activated, the ADB class 
redirects rotation requests to the Arduino class. The Arduino, in turn, controls the positioning of the smartphone through serial communication using the PySerial library, thus enabling specific guidance during tests.

\subsection{Research Questions in the Replication Study}
\label{sec:research_questions_replication}

The empirical evaluation of DLD was conducted using a virtual device (AVD). Therefore, it did not suffer any influence from hardware sensors, connection technologies, memory constraints, and processing power. 
Our conceptual replication study was carried out in a different context: we had the support of the \tname{} robotic infrastructure and physical rotations of a real device in order to promote less invasive and more realistic testing.

DLD was also compared DLD with ALARic~\cite{alaric} and QUANTUM~\cite{quantum}. Since our goal is not to make comparisons against competitors that had already been compared against DLD, this analysis was discarded. The third question in the original experiment was about the effectiveness of DLD, which led to our first Research Question (RQ1). The fourth question was about the discovery of bugs by the two types of oracles, which led us to our RQ2. Finally, the fifth question was about the relevance of data loss for developers, leading us to our RQ3.
Thus, the following research questions were defined for our conceptual replication study.

\begin{itemize}
    \item [\textbf{RQ1}] What is the relationship between True Bugs and False Alarms identified by the \tname{} infrastructure?
    \item [\textbf{RQ2}] Does the detection of data loss follow the same proportion identified by DLD for the \textit{screenshot-based} and \textit{property-based} oracles?
    \item [\textbf{RQ3}] Is data loss (still) relevant to developers?
\end{itemize}
RQ1 investigates whether the proportion of True Bugs and False Alarms identified by \tname{} is similar to that observed in the DLD evaluation study~\cite{DLD}. 
RQ2, on its turn, help us to clarify whether the two oracles behave similarly to what was observed in the DLD evaluation study~\cite{DLD} (taking into consideration that we used different apps, a real device and physical orientation changes).
And RQ3 aims to gather information from developers regarding reported data loss flaws and to determine whether there is still an interest in addressing them.


\section{Methodology}
\label{sec:methodology}

The aim of this experiment is to investigate data loss occurrences across a broader range of Android applications in order to report and propose solutions to developers. The Android store chosen for this investigation was \mbox{F-Droid}, as it hosts over 2,500 free and open-source projects, along with providing a link to download the compiled version of the application. 

The F-Droid store offers 17 categories of applications like Development, Writing, Internet, Security, System, and so on. Some applications belong to multiple categories; for instance, the EteSync app falls under Writing, Internet, Security, and System. Our subjects encompasses all 17 categories available in the F-Droid store.


\subsection{Data Collection and Subject Selection}

The data collection process was performed using web scraping techniques to obtain the name, the description, the Android version, and the links to the repository and the compiled version. Initially, all links in the 17 software categories were collected, totaling 3,631 entries. After manually removing duplicated entries and broken links, this number was reduced to 2,461.

The next step was to perform an analysis of the Manifest of the applications to check for any screen orientation restrictions. The analysis employed a library to read the AndroidManifest.xml file integrated into the compiled version of the application in order to look for specific configuration instructions.

The Androguard library was used to search for occurrences of the \textit{screenOrientation} instruction in the application activities. Three types of restrictions may occur: (i) complete blocking of the application orientation, in which case \tname{} cannot trigger the stop-start event (thus, we discarded such apps); (ii) partial blocking of orientation changes in the application, in which some activities may block a rotation (as \tname{} cannot differentiate between a blocked and a normal activity, which potentially leads it to a prolonged exploration without being able to execute the stop-start event, we set such apps to a lower priority for selection); and (iii) applications with no orientation change restrictions, which were assigned a higher priority for selection.

Another instruction sought was \textit{configChanges}, which is used to assign the responsibility for configuration changes to the developer~\cite{AndroidConfigChanges}. This is done through the implementation of the onConfigurationChanged() method in order to handle configuration changes. In practice, it is common for developers not to implement this method but to solely use this instruction to prevent the destruction of the activity during orientation changes, which can lead to side effects.

An analysis of the Manifest revealed that 30.8\% of the applications have only one activity. This number increases to 55.6\% when considering applications with up to 3 activities, and to 70.5\% when considering applications with up to 5 activities. Regarding configuration changes, 35.1\% of the projects have the configChanges parameter enabled in the Manifest, indicating that the developer assumes responsibility for handling events such as orientation changes via the onConfigurationChanged() method. Additionally, 22.8\% of the applications have some form of orientation lock. This can occur in all activities or only in some of them. When considering only locks in the main activity, this percentage decreases to 13.8\%. As most of the application functionality is generally found in the main activity, the presence of this lock automatically disqualifies the application. However, when it occurs elsewhere, the application may be analyzed at a later time. The screenOrientation criterion was used to eliminate applications that had locks in any activity. Thus, the number of eligible projects was reduced to 1,899.

The next criterion aims to prevent choosing
abandoned projects. We selected apps that had at least one update in the year before the experiment.
When analyzing the repositories used by F-Droid, it was observed that 78.4\% correspond to GitHub, 11.3\% to GitLab, and 10.3\% to other repositories. A script to obtain the commits for each project was developed for GitHub to maximize the number of evaluated applications, while the other repositories were discarded. By applying these restrictions, \totalNumAppsCandidate{} projects were selected, which corresponds to 29.8\% of the initial number of projects.

\subsection{Execution}

The applications selected in the previous phase were subjected to \tname{}. The experiment employed the same number of events (2,250 per app) as used in the DLD evaluation, except that these events were carried out with real rotations, aiming to uncover genuine defects in a larger set of applications.

The execution process was automated to run continuously, respecting the time required for installation and uninstallation on the Android smartphone. A total of \totalNumAppsAnalyzed{} applications were randomly selected for analysis in the experiment (approximately 10\% of the number of candidate applications). In the experiment, 36 applications required a minimum of Android 4, 27 applications required at least Android 5, and 14 applications required Android 6.

The experiment was conducted using \tname{} with a Motorola Moto G30 smartphone (4GB memory, 128GB storage, 1,600$\times$720 resolution, and Android 11). Initially, some device settings were configured to allow application installation and USB communication. Additionally, a \textit{Null Keyboard} was installed to prevent the keyboard from appearing during text input. This decision was made based on the observation of False Alarms related to keyboard interference during our initial analysis.

Upon initiating the experiment with 2,250 events per app, we observed that the processing time was 50\% longer (4.5 hours) than the time reported in the DLD study, which took 3 hours. This was due to the difference between the simulator (AVD) and the real device, and was not related to the use of the real rotation done by the robot. In other words, if the DLD experiment had been executed on a real smartphone (with no robot) instead of an AVD, the time required to process the 2,250 events would also have been 50\% longer (4.5 hours).

The experiment generated \totalNumAppsAnalyzed{} reports, with an average of 192 rotations, 82 data losses, 71\% coverage of the activities, and 18.5\% fatal exceptions per application. 
The individual reports, one per app, are available in our replication package (see Section~\ref{sec:data}).
The average execution time for each run was 4.5 hours, totaling approximately \totalhoras{} hours of computational processing.

\subsection{Report Evaluation}

We manually reviewed the information in the reports to categorize alarms as either true positives (True Bug) or false positives (False Alarm). A false alarm happens when the second screen capture shows expected behavior of the application, such as an animated element, a blinking cursor, timers, timed messages, etc.
All data losses marked as True Bugs were reported to the developers by opening issues with descriptions of the problems using the artifacts produced by \tname{}.

False Alarms were more common in data losses reported via screenshots and were associated with animations, videos, or timed messages (toasts). They can also occur due to differences in screenshot shading resulting from delayed screen updates or due to application closure caused by an exception.

A data loss is classified as a True Bug when a variable is destroyed, reset, or assigned a default value. When reported via screenshots, data loss manifests as disappearing fragments, text, or messages, elements that change color or lose their state, among other manifestations. These losses are always associated with significant differences between the views files.

\subsection{Reporting Issues}

After classifying the alerts, we reported the True Bugs to the developers by opening an issue that includes a problem description, steps to reproduce, expected behavior, device model, and application version. The screenshots reported by \tname{} were used to facilitate the procedures for the developer to reproduce the issue.

\section{Analysis and Discussion of Results}
\label{sec:analysis}

In this section, we describe the analysis of the results for each research question.

\subsection*{RQ1: True Bugs and False Alarms identified by the \tname{} infrastructure}
\label{RQ1.1}
In our replication study we analyzed \totalNumAppsAnalyzed{} applications using \tname{}, with an average execution time of 4.5 hours
per application, and evaluated the \totalNumWarnings{} data loss alerts generated. Six of these applications did not report any bug alerts, as shown in Table~\ref{tab_withoutbugs}. In three cases, \tname{} got stuck on the tutorial screen, and to overcome this an initial configuration would be needed to advance to the main activity before starting the evaluation. Six applications reported bug alerts, though all of them were subsequently classified as False Alarms (see Table~\ref{tab_withoutbugs_false}).

\begin{table}[htbp]
    \caption{Applications that did not produce alerts}
    \label{tab_withoutbugs}
    \begin{center}
    \setlength{\tabcolsep}{4pt}
    \begin{tabularx}{\columnwidth}{@{}lX@{}}
        \toprule
             \textbf{Application} & \textbf{Diagnosis}\\
             \midrule
             TripleCamel & 
             Simple interface with only usage instructions\\
             AF Weather & Widget\\
             Shelter & Tutorial prevented access to the main activity\\
             Track and Graph & Tutorial prevented access to the main activity\\
             Keymapper & Tutorial prevented access to the main activity\\
             AccA & The application requires \textit{root} access to function\\
            \bottomrule
        \end{tabularx}
    \end{center}
\end{table}

\begin{table}[htbp]
\centering
\caption{Applications that generated False Alarms only}
\label{tab_withoutbugs_false}
\setlength{\tabcolsep}{3pt}
\begin{tabularx}{8.5cm}{@{}lrrX@{}}
\toprule
\textbf{Application}   & \multicolumn{1}{l}{\textbf{\#Alerts}} & \multicolumn{1}{l}{\textbf{\#True Bugs}} & \textbf{Diagnosis} \\ \midrule
Compass               & 1                                    & 0                                           & Update                         \\
Termux:Boot           & 1                                    & 0                                           & Capture error                     \\
Egyptian Mouse Pounce & 163                                  & 0                                           & Animation                            \\
DSA Assistant         & 6                                    & 0                                           & Timed message                \\ \bottomrule
\end{tabularx}
\end{table}

\begin{figure} 
    \begin{subfloat}{}
        \centering
        \begin{tikzpicture}
        \begin{axis}[
            width=\columnwidth,
            height=2.5cm,
            boxplot/draw direction=x,
            ytick={1},
            yticklabels={TB\label{graphics_true_and_false_bug_tb}},
            ]
            \addplot+[
                boxplot prepared={
                    median=17,
                    upper quartile=35,
                    lower quartile=4,
                    upper whisker=84,
                    lower whisker=2,
                    box extend=0.4,
                },
            ] coordinates {};
            
            \addplot+[scatter, only marks, mark options={fill=white, draw=black}]coordinates{
                (90, 1)
                (102, 1)
                (123, 1)
                (158, 1)
                (321, 1)
            };
        \end{axis}
    \end{tikzpicture}
        
    \end{subfloat}
    
    \begin{subfloat}{}
        \centering
        \begin{tikzpicture}
        \begin{axis}[
            width=\columnwidth,
            height=2.5cm,
            boxplot/draw direction=x,
            xmode=log, 
            ytick={1},
            yticklabels={FA\label{graphics_true_and_false_bug_fa}},
        ]
            \addplot+[
                boxplot prepared={
                    median=1,
                    upper quartile=6,
                    lower quartile=0.8,
                    upper whisker=15,
                    lower whisker=0.8,
                    box extend=0.4,
                },
            ] coordinates {};
            
            \addplot+[scatter, only marks, mark options={fill=white, draw=black}] coordinates {
                (17, 1)
                (28, 1)
                (29 1)
                (35, 1)                
                (96, 1)
                (163, 1)
                (219, 1)
                (321, 1)                
                (379, 1)
            };
        \end{axis}
    \end{tikzpicture}
        
    \end{subfloat}

   \caption{Distribution of True Bugs and False Alarms \label{graphics_true_and_false_bug}}
\end{figure}
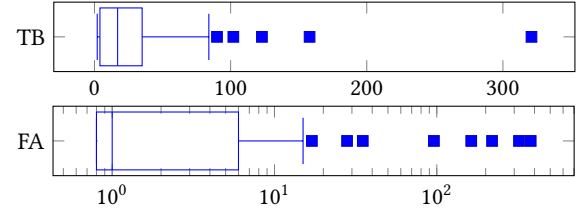

After a manual analysis of each alert, \totalNumTrueBugs{} (\percentualDeTrueBugsTotal{}) alerts were classified as True Bugs, and \totalNumFalseAlarm{} (\percentualDeFalseAlarmTotal{}) were classified as False Alarms. Figure~\ref{graphics_true_and_false_bug} presents box plots of these data. The $x$-axis represents the quantity of True Bugs and False Alarms, respectively, while the points represent the applications.

Upon analyzing the reasons for the number of False Alarms, we observed that nine applications exhibited outliers in their data. 
While most applications had up to ten False Alarms, five applications fell within the range of ten to one hundred alarms, and the remaining four had over one hundred False Alarms. A more detailed investigation revealed that the four applications generating an excessive number of False Alarms (totalling \totalCincoAppAnimationFalseAlarm{} False Alarms and  \totalCincoAppAnimationTrueBugs{} True Bugs) were applications that had animations, which confuse the screenshot oracle. Table~\ref{tab_outliers} presents the number of True Bugs, False Alarms, and the diagnosis of the reasons for False Alarms for the considered outlier applications. Although the Moonlight application (line 6 in Table~\ref{tab_outliers}) has animations, they only occur when data is loaded. These False Alarms could possibly be reduced by adjusting the time between the before and after state captures.

\begin{table}[htbp]
    \caption{Analysis of outlier observations}
    \label{tab_outliers}
    \begin{center}
        \begin{tabular}{lrrl}
        \toprule
             \textbf{Application} & \textbf{TB}& \textbf{FA} & \textbf{Diagnosis}\\
             \midrule
             Material Files & 97 & 28 & Blinking cursor\\
             Wikipedia & 35 & 28 & Keyboard rendering\\
             VocableTrainer & 20 & 28 & Keyboard rendering\\
             A Time Tracker & 49 & 17 & Time change\\
             Goodtime & 6 & 312 & Clock animation\\
             Moonlight & 45 & 98 & Data loading animation\\ 
             OpenPods & 0 & 163 & Animation during app usage\\
             Baby Dots & 1 & 379 & App animation\\
             Fiddle Assistant & 18 & 219 & App animation\\
            \bottomrule
        \end{tabular}
    \end{center}
\begin{flushleft}
\scriptsize
\textit{TB: True Bugs; FA: False Alarms}
\end{flushleft}
\end{table}

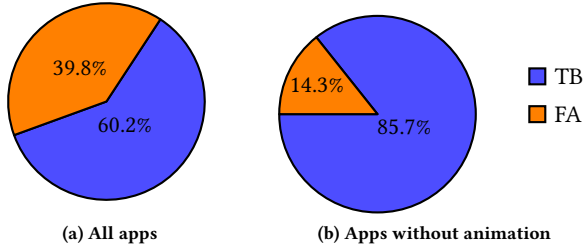
\begin{figure}[htb]
    \subfloat[All apps]{
    \begin{tikzpicture}
        \pie[radius=1.3, rotate=200, color={blue!70, orange!100!black}]    
        {60.2/, 39.8/} 
    \end{tikzpicture}    
    }%
    \qquad
    \subfloat[Apps without animation]{
    \begin{tikzpicture}
        \pie[radius=1.3, rotate=180, color={blue!70, orange!100!black}, text=legend]    
        {85.7/TB, 14.3/FA} 
    \end{tikzpicture}    
    }%
    \caption{True Bugs (TB) and False Alarms (FA) before and after excluding apps with animations.
\label{graphics_classificacao}}
\end{figure}

When considering all applications, including those with animations, we reach the proportion of \proporcaoTrueBugGeral{} True Bugs for each False Alarm. However, when we exclude the applications with 100+ False Alarms
(i.e., Goodtime, OpenPods, Baby Dots, and Fiddle Assistant), 
the proportion rises to \proporcaoTrueBugWithoutAnima{} True Bugs for each False Alarm. Figure~\ref{graphics_classificacao} presents this difference for both cases.

The coverage percentages of activities during the experiment were also analyzed. Figure~\ref{graphics_covered} presents the box plots, where the $x$-axis represents the percentage of activity coverage, the points represent the applications.
Most applications had coverage above \coberturaPriQuartil{}, with a median of approximately \coberturaMediana. Even applications with coverage below \coberturaPriQuartil{} managed to find a non-negligible number of True Bugs, with a total of \coberturaTrueBug{} True Bugs. The number of False Alarms for low-coverage applications was \coberturaFalseAlarm. However, this high number was caused by the \coberturaAppOutlier{} application, which presented \coberturaFalseAlarmOutlier{} False Alarms and revealed only \coberturaTrueBugOutlier{} True Bug.

\begin{figure}[tb]
        \centering
        \begin{tikzpicture}
        \begin{axis}
            [
            width=\columnwidth,
            height=2.5cm,
            boxplot/draw direction=x,
            ytick={1},
            yticklabel style={text width=1.4cm, align=center},
            yticklabels={Double Rotation},
            ]
            \addplot+[
                boxplot prepared={
                    median=191,
                    upper quartile=223,
                    lower quartile=144,
                    upper whisker=380,
                    lower whisker=8,
                    box extend=0.4,
                },
            ] coordinates {};
            
            \addplot+[scatter, only marks, mark options={fill=white, draw=black}] coordinates {
                (0,1)
                (380, 1)
                (566, 1)
            };
        \end{axis}
        \end{tikzpicture}
    \caption{Oracle checks per app.}
    \label{graphics_oraculo}
\end{figure}
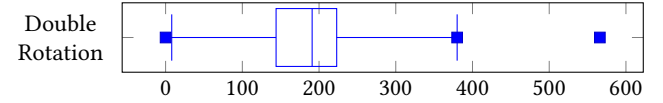

Another point analyzed was the number of oracle checks indicated by the double rotation in the generated report, Figure~\ref{graphics_oraculo}. Screen rotation is the stop-start event with the highest probability of revealing data loss issues due to the destruction and recreation of the activity.
Five applications behaved as outliers: two below the lower limit and three above the upper limit of the distribution. Most applications had between 145 and 230 oracle checks, with a median of 200 checks.

A more detailed analysis of the results of these applications confirmed the intuition that the number of checks performed is not necessarily correlated with the number of True Bugs revealed. For some applications, the oracle can be applied hundreds of times and generate few alerts, but they correspond to True Bugs; for other applications, hundreds of alerts can be generated, but all are related to False Alarms. Table~\ref{tab_oraculo} presents the number of alerts, True Bugs, and False Alarms generated for the outlier applications based on the number of checks performed.

The variation in the number of double rotations among applications is related to the DLD exploration strategy of interacting with all elements of the activity to insert values different from the default values.
In some cases, the \tname{} can get stuck in an activity if it requires a very specific sequence of screen touches to change activities. For example, in the Babydots application, if a touch event is launched at a specific point on the screen, the application is locked, requiring a sequence of actions to unlock it. In these cases, \tname{} waits for a condition (for example, inserting different values from the default value) that ends up consuming many events.

\begin{figure}[tb]
        \centering
        \begin{tikzpicture}
        \begin{axis}
            [
            width=\columnwidth,
            height=2.5cm,
            boxplot/draw direction=x,
            ytick={1},
            yticklabel style={text width=1.4cm, align=center},
            yticklabels={Activity Coverage},
            ]
            \addplot+[
                boxplot prepared={
                    median=66.5,
                    upper quartile=100,
                    lower quartile=40,
                    upper whisker=100,
                    lower whisker=9
                },
            ] coordinates {};
        \end{axis}
        \end{tikzpicture}
    \caption{Activity coverage per app.}
    \label{graphics_covered}
\end{figure}
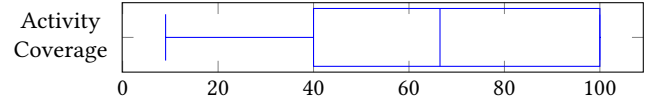

\begin{table}[htb]
    \caption{Outliers in data loss analysis due to double rotation. }
    \label{tab_oraculo}
\begin{tabular}{lrrrr}
\toprule
\textbf{Projects} & \textbf{DR} & \textbf{AL} & \textbf{TB} & \textbf{FA} \\ 
\midrule
Bubble            & 8                                    & 3                              & 3                   & 0 \\ 
Step-and-height-counter & 12                                   & 12                             & 12                  & 0 \\ 
Baby dots      & 380                                  & 380                            & 1                   & 379 \\ 
Compass            & 566                                  & 1                              & 0                   & 1 \\ 
Mindustry          & 566                                  & 1                              & 1                   & 0   \\ 
\bottomrule
\end{tabular}
\begin{flushleft}
\scriptsize
\textit{DR: Double Rotation; AL: Alerts; TB: True Bugs; FA: False Alarms}
\end{flushleft}
\end{table}

\begin{tcolorbox}[top=4pt,left=2pt,right=2pt,bottom=4pt]
\textbf{Response to RQ1:} When considering all alerts, the \tname{} achieved \proporcaoTrueBugGeral{} True Bugs for each False Alarm. However, when removing applications with animations, the ratio increases to \proporcaoTrueBugWithoutAnima{} True Bugs for each False Alarm. To put into perspective, the ratio in the DLD study was 4 True Bugs to each False Alarm. 
Such a difference could be related to the type of applications evaluated, as the removal of five applications with animations caused a significant change in this ratio.
\end{tcolorbox}

\subsection*{RQ2: Comparison with DLD in terms of oracles}
\label{RQ1.2}

The empirical evaluation of DLD~\cite{DLD} used both screenshot-based and property-based oracles. A bug alert can be triggered by one or both of these oracles. The authors of DLD manually classified the alerts into True Bugs and False Alarms and reported that 73.1\% of True Bugs were identified by both oracles, 17.8\% were identified only by the property-based oracle, and 9.2\% were identified only by the screenshot-based oracle. Regarding False Alarms, 73.6\% occurred in both oracles, 21.1\% in the screenshot-based oracles, and 0.1\% in the property-based oracles. In 5.3\% of cases, the oracle failed due to errors in capturing correct information caused by delays in app updates. As a solution, the authors suggested adjusting the parameters of DLD to reduce these failures.

\subsubsection*{Analysis of the experiment with all applications}
\begin{figure}
    \subfloat[True Bugs]{
    \begin{tikzpicture}
        \pie[radius=1.3, rotate=90, color={blue!70, orange!100!black, red!70}]    
        {71.9/, 9.7/, 18.4/} 
    \end{tikzpicture}    
    }%
    \qquad
    \subfloat[False Alarms]{
\begin{tikzpicture}
        \pie[radius=1.3, rotate=60, color={blue!70, orange!100!black, red!70}, text=legend]    
        {45.2/BH, 52.8/SBO, 2/PBO} 
        \draw[rotate=55, <-, thick] (1.4,0) -- (1.6,0) node[left] at (2,-0.3) {\large 2\%};
    \end{tikzpicture}
    
    }%
    \begin{flushleft}
    \scriptsize
    \textit{BH: Both Oracle, SBO: Screenshot-based Oracle, PBO: Property-based Oracle}
    \end{flushleft}
    \caption{True Bugs and False Alarms for the evaluated apps}
    \label{graphics_oracle_true_bug}    
\end{figure}
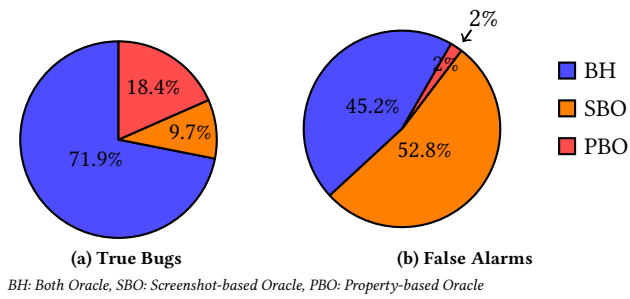

Considering all applications, without excluding those with animations, the results observed in our study were similar to the results reported by the authors of DLD for True Bugs. However, when compared to False Alarms, there is a different distribution of data.
Figures~\ref{graphics_oracle_true_bug}(a) and \ref{graphics_oracle_true_bug}(b) present the distribution of oracles related to True Bugs and False Alarms, respectively, considering all applications.
Regarding True Bugs, both oracles were able to identify data loss in \perctBothTotalTrueBug{} of cases. For the property-based oracle, this value was \perctPropertyTotalTrueBug{}, and for the screenshot-based oracle, \perctSnapTotalTrueBug. In the DLD evaluation, the results were 73.1\%, 17.8\%, and 9.2\%, respectively.
Concerning False Alarms, Figure~\ref{graphics_oracle_true_bug}(b) shows that both oracles were able to identify data loss in \perctBothTotalFalseAlarm{} of cases. For the property-based oracle, this value was \perctPropertyTotalFalseAlarm{}, and for the screenshot-based oracle, \perctSnapTotalFalseAlarm. In the DLD evaluation, the results were 73.6\%, 0.1\%, and 21.1\%, respectively.

In the case of False Alarms, there is an additional category related to information capture errors, accounting for 5.3\% in the DLD study. This error is associated with the slowness of the application in rendering the activity, which leads the oracle to detect differences between captures due to outdated information. This problem can be reduced by adjusting the time parameter for the oracle.
For the purpose of comparison, False Alarms related to screen update time were attributed to this category because when the oracle capture occurs before the expected time, differences in tones or incomplete rotations between the images may occur.

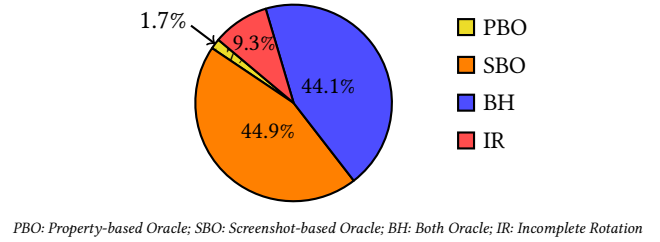
\begin{figure}
    \centering
    \begin{tikzpicture}
        \pie[radius=1.3, rotate=140, color={yellow!90!black, orange!100!black, blue!70, red!70}, text=legend]    
        {1.7/PBO, 44.9/SBO, 44.1/BH, 9.3/IR}        
        \draw[rotate=143, <-, thick] (1.3,0) -- (1.7,0) node[left] at (1.7,-0.1) {\large 1.7\%};
    \end{tikzpicture}   
    \begin{flushleft}
        \scriptsize
        \textit{PBO: Property-based Oracle; SBO: Screenshot-based Oracle; BH: Both Oracle; IR: Incomplete Rotation}
    \end{flushleft}
    \caption{Proportion of False Alarms separated by oracle and the criterion of incomplete rotation}
    \label{graphics_falsoAlarmeComRotacao}
\end{figure}

Figure~\ref{graphics_falsoAlarmeComRotacao} illustrates the distribution of False Alarms into four categories. Both oracles made mistakes together in \perctBothTotalFalseAlarmRotation{} of cases, while the property-based oracle made errors in \perctPropertyTotalFalseAlarmRotation{} of cases, and the screenshot-based oracle made errors in \perctSnapTotalFalseAlarmRotation{} of cases. Incomplete rotations accounted for \perctTotalFalseAlarmRotationIncompleta{} of cases. 
Incomplete rotation occurs when the screenshot capture takes place before the orientation change is completed, and it may be related to the smartphone processing speed at the time of capture. 
The distribution of oracles concerning False Alarms exhibited a different behavior than that observed in the DLD study, both in terms of quantity and proportion. 
However, the property-based oracle had fewer failures in both cases. The observed differences are related to the characteristics of the evaluated applications.

\subsubsection*{Differences between the applications in both studies (DLD and \tname{})}
In our conceptual replication study, cases of True Bugs reported with and without animation exhibited a similar behavior to the original study. However, there was a significant difference regarding False Alarms, both in quantity and proportion.
The observed differences may be related to the characteristics of the analyzed applications, especially in cases of applications with animations. For example, applications that load data from the internet may require more time before calling the oracles in comparison to the default time (our study did not adjust the time parameter between events). %

The large number of False Alarms does not necessarily implies a higher manual effort during the alert analysis phase. As this type of alert usually occurs in a specific activity
and is detected by only one oracle, it is possible to apply filters to the generated alerts and group the captures for more efficient manual or automated analysis.

\begin{tcolorbox}[top=4pt,left=2pt,right=2pt,bottom=4pt]
\textbf{Response to RQ2:} 
When considering all applications, True Bugs had a distribution similar to that observed in the DLD study. 
The distribution remained similar when conducting the same analysis after removing applications with animations. 
However, we observed differences in False Alarms when comparing with the DLD study. 
\tname{} obtained a higher proportion of False Alarms in the screenshot-based oracle when considering the grouping of applications with and without animation. This difference may be related to the characteristics of the analyzed applications.
\end{tcolorbox}

\subsection*{RQ3: Relevance of data loss bugs to developers}
\label{RQ1.3}

After manually analyzing all \totalNumWarnings{} warnings and classifying them as True Bugs and False Alarms, we found that many warnings were duplicated, sometimes reporting the same failure identically, and in other cases, with minor variations. For example, a form containing the string "test" and the same form containing "test123" reported different warnings. Grouping similar failures resulted in a total of \totalBugsWithoutDuplication{} bugs. Only 5 failures were not reported because they were associated with archived projects (Bubble and TrebleShot). The remaining \totalBugsWithoutDuplicationNotArchived{} failures found were reported to developers by opening issues on the project repositories on GitHub.
Due to the quantity of bugs per project, one or more issues were created to report them, always indicating the relationship with the others, resulting in a total of \issueTotal{} issues opened.

To provide context to developers regarding the data loss issue, the issue descriptions included concepts of the activity lifecycle, data loss problem, \tname{} detection technique, and possible solutions described in the Android documentation. After this contextualization, for each bug, the steps to reproduce the bug, the expected result, the configuration environment, and some links to Android documentation were reported. Additionally, some artifacts produced during the \tname{} execution, such as the screenshots before and after the double rotation, the changed properties, and the touchscreen taps, were included.

At the time of writing, \issueAnswered{} issues have been addressed (\issueAnsweredPercentage{} of the reported issues), and \issueAccepted{} (\issueAnsweredAcceptedPercentage{} of the addressed issues) received positive feedback, demonstrating developers interest in the problem. A detailed breakdown of the status of each issue is available in our replication package.
Among the issues with negative feedback (only \issueRejected{}), developers reported no interest in resolving the issue because it was a very specific problem with small probability of being reproduced by a user. Others attributed the problem to Android or reported that the project is no longer maintained or simply closed the issue without responding.

For issues with positive feedback, \issueAcceptedFix{} were resolved by project members. In some cases, a partial solution was developed promptly, especially when the reported failure compromised an essential application feature. In one particular case, the developer was about to release a newly restructured version of the application, and the bugs had already been fixed almost simultaneously with the issue opening. In \issueAcceptedOpen{} cases, the developers left the issue open, waiting for a convenient opportunity to resolve them. In \issueAcceptedWontfix{} cases, the developers acknowledged the relevance of the data loss problem but expressed no interest in resolving it and left the issue open. 

\begin{center}
\begin{tcolorbox}[top=4pt,left=2pt,right=2pt,bottom=4pt]
\textbf{Response to RQ3:} A total of \totalBugsWithoutDuplication{} bugs were reported in \issueTotal{} issues, with \issueAnswered{} (\issueAnsweredPercentage) issues receiving responses from the developers. Among those addressed, \issueAccepted{} issues had data loss bugs confirmed. These accepted issues correspond to \issuefalhasConfirmadas{} bugs, representing \issueAnsweredAcceptedPercentage{} of the addressed bugs. Thus, it can be inferred that the data loss problem is relevant to developers.
\end{tcolorbox}
\end{center}

\section{Feasibility Study of Robotic Automation for Data Loss Testing}
\label{sec:feasibility}

\myparagraphtwo{What are the costs of setting up the testing environment?} In a context where the team already works with robots, incorporating data loss testing is a straightforward task because the effort is in integrating DLD with the robot control system. However, when it is necessary to build the entire infrastructure, the implementation cost can vary significantly. If the team chooses to acquire an industrial robot capable of performing various types of movements with different degrees of freedom and speed control, this cost can reach tens of thousands of dollars. The handmade robotic arm built as part of this work to rotate the smartphone has a much lower cost (the version presented in Figure~\ref{fig:robo_esquema}(a) costs approximately \$60). During the replication experiment, which lasted approximately 350 hours, the robotic arm did not present defects or require additional calibrations due to the number of orientation changes. The calibration is performed when the equipment is turned on, and no further calibration is required during testing.

\myparagraphtwo{What are the challenges in integrating the robotic approach into a data loss detection tool?} Integrating the automated testing tool with the robot is a relatively simple task that involves intercepting ADB commands and replacing them with robotic instructions. In this case, the robotic instruction is a command to rotate an object attached to a support. However, this depends on an API that the tool provides or on its source code availability. The integration with DLD was possible because the developers provided the source code. Another critical point is the speed of the robotic arm in performing an orientation change. During the validation experiment of the robotic arm, we verified that this time was less than 2 seconds for each orientation change. By default, DLD uses a 3-second delay between events, allowing the robotic arm to perform both orientation changes without interfering with the oracle verification.

\myparagraphtwo{Are there advantages to using physical devices compared to emulated ones?} An AVD (Android Virtual Device) is a simulation of a hardware profile to run on the Android emulator, where the components are abstractions of real devices. Although it resembles a physical device, an AVD has limitations. For example, it is not possible to run tests involving phone calls, Bluetooth connections, or image rectification algorithms (which require the use of the camera).

\section{Threats to Validity}
\label{sec:threats}

The primary internal threat to our study is the manual classification of alerts as either True Bugs or False Alarms. Although this classification is a straightforward process involving the verification of differences in properties or screenshots, it is not a trivial task. It needs an understanding of the application behavior to determine whether a supposedly data loss is, in fact, an expected behavior of the application. For example, in cases where the activity involves a timer, the oracles for screenshots, properties, or both can indicate (erroneously) a data loss.

The primary external threat pertains to the generalizability of the results. Our study was conducted on a sample of \totalNumAppsAnalyzed{} applications collected from an Android app store. Mitigating this threat can only be achieved through additional studies that consider different applications, domains, and additional research questions.

The choice of a single smartphone model and a specific version of Android was made to maintain control and consistency throughout the experiment. This allowed us to focus on identifying issues and observing app behavior in a standardized environment. However, we acknowledge that the results may not be generalizable to other devices or Android versions, underscoring the need for future research that includes a greater variability of hardware and different operating system configurations to achieve a more comprehensive evaluation.

\section{Related Work}
\label{sec:related}

\subsection{Testing for data loss detection}

Zaeen \etal{}~\cite{quantum} introduce QUANTUM, an automatic test case generator exploring GUIs in mobile apps. They selected 106 reproducible bugs from 13 applications to identify opportunities for automatic test case generation. Each bug was manually categorized to identify the type of oracle that could be used and grouped into categories. 
It generated 60 tests, unveiling 22 bugs, emphasizing GUI failures and also addressing data loss during the activity lifecycle exploration.

Amalfitano \etal{}~\cite{orientation} conducted two blackbox experiments with a non-invasive approach, aiming to identify GUI failures (data losses) through orientation changes on real devices. 
The first experiment 
found 439 GUI failures among 68 applications.
The second experiment 
assessed interface failure in large company applications. Ten applications were evaluated and revealed 140 flaws.
Both experiments involved manual detection, checking for screen changes after double rotation. 

Riccio \etal{}~\cite{alaric} present ALARic, a tool that uses dynamic exploration techniques over particular events like double orientation change and semi-transparent activity intent. 
ALARic tested 15 apps and revealed 106 True Bugs, emphasizing the efficacy of double orientation in exposing an activity lifecycle issue.

As already described at Section~\ref{sec:dld}, Riganelli \etal{} \cite{DLD} developed the Data Loss Detector (DLD)  
to identify data loss on Android devices using
ADB (Android Debug Bridge) commands to simulate device interaction. 

The main benefits of using DLD were the blackbox approach and the automatic test generation, which employed an exploratory strategy to cover the activities and an interaction strategy to change default element values. 

Guo \etal{}~\cite{DetectingFixDataloss, iFixDataloss} present iFixDataloss, a tool combining static and dynamic analysis for data loss detection and correction. For each application's activity, predefined tests and events covering various data loss scenarios are executed. When a data loss problem is encountered, 
the tool uses a model to generate a correction patch. After creating the patches for each variable, a new verification is performed on the corrected activities to check for any remaining data loss problems or if the corrections caused any crashes.

The works of Amalfitano and Riganelli present non-invasive blackbox approaches for data loss detection, one manual and the other automated. Guo's work (iFixDataloss) introduces a greybox approach, incorporating static code analysis into a systematic, automated, and invasive exploration. \tname\ adds value by leveraging robotics for less invasive blackbox testing, physically performing a smartphone rotation.


\subsection{Use of robots for smartphone testing}

Robots are widely used for many repetitive tasks, and interest in using them in the context of mobile device testing has been growing recently~\cite{Maciel2022smsRoboticTesting}.
Banerjee \etal{}~\cite{JitterReduction, FaceAuthentication, RecognitionFace, ImageRectification, rastreamento} employed robotic testing to simulate real-world user movements. This methodology proved instrumental in algorithm comparison across diverse user scenarios, mitigating the influence of human errors. The use of robots ensures a controlled environment for systematic algorithm evaluation, reduced manual testing, and improved efficiency and scalability.

K. Mao \etal{}~\cite{AxizBlackBoxAutomation} presented the Axiz tool, a robotic test generator for mobile applications using a real device in a black-box testing approach. Common points between Axiz and \tname{} include the use of a robot to execute tests, the utilization of the device cyber-physical interface, and the automatic test generation. However, in order to create efficient test cases, the Axiz tool requires a predefined test suite to generate a realistic model. This may lead to two issues: biases can be introduced by the lack of test diversity, and bias can be introduced by the testing team programming tendencies. \tname{} generates realistic tests without the need to access a test suite, as it uses DLD to identify activity elements and their possible combinations, generating random events that adhere to these requirements.

Qian \etal{}~\cite{RoScript} presented the RoScript tool, a script-based robotic testing system employing a non-intrusive approach for GUI testing on touch-screen devices.
Unlike intrusive approaches that obtain GUI information through access to the device operating system, RoScript uses computer vision to identify GUI elements and states. It interacts with the device using the cyber-physical interface. This process differs from traditional approaches that use installed apps on the device under test to record these steps.

Craciunescu \etal{}~\cite{RobotDedicated} introduced a portable, low-cost robot developed for testing mobile devices with resistive and capacitive touch screens. By employing image processing techniques, it can provide feedback on executed actions and remains robust in locating elements on the screen even with device repositioning. 

Juang and Cheng~\cite{characterRecognition} employed computer vision and a robotic arm to conduct tests on smartphones with the aim of creating a robotic system that accepts a test case and executes actions similar to a human on a device.

Vermaa \etal{}~\cite{MultiTouchArduino} presented a low-cost robot that allows for actions on a smartphone using the touch screen, constructed with a Cartesian coordinate design using an Arduino board, two Nema 17 stepper motors, and some other components.

\section{Conclusion and Future Work}
\label{sec:conclusion}

Our work proposed an automated testing framework, activated by a robot, to perform data loss tests in a real environment. This proposal is more realistic and less invasive than recent previous work~\cite{alaric, DLD, iFixDataloss}. It is more realistic because the orientation changes are triggered by the physical rotation of the smartphone, thus interacting with the sensors as in a real usage scenario. It is less invasive because a crucial portion of the interaction with the smartphone (the orientation change) is carried out by a low-cost homemade robot. In other words, these interactions are not sent through ADB commands via the USB port to the smartphone.

In order to evaluate our proposal, we conducted an empirical assessment with \totalNumAppsAnalyzed{} randomly selected Android apps from an app store and identified \totalBugsWithoutDuplication{} data loss bugs. We reported all bugs to the developers and received responses to \issueAnsweredPercentage{} of them.  The developers acknowledged the problem in \issueAnsweredAcceptedPercentage{} of the cases. 


As future work, we aim to investigate the feasibility of proposing a completely non-invasive approach~\cite{Freitas_RFNIT_FSE24}. We plan to replace the screenshot oracle with computer vision techniques to identify differences between activities after a double rotation event. 
We also plan to investigate other ways to trigger a stop-start event to detect data loss, such as switching between apps under stress conditions associated with increased processing. 
Another area we aim to improve is the activity exploration strategy. The current solution does not allow for selecting the activities we want to test, such as choosing a specific activity, excluding those with blocked orientation changes, among others. During app selection, it was observed that many of them have orientation locks in one or more activities. 
It is also possible to enhance the way we classify alerts into True Bugs and False Alarms using computational intelligence techniques following the example of previous works~\cite{miranda2020prioritizing,cabral2022rvprio}. 
One potential avenue of future work is investigating the feasibility of using binary classification algorithms to categorize the alerts generated by \tname{} into True Bugs or False Alarms.


\section{Data Availability}
\label{sec:data}
To facilitate independent verification and replication, our artifacts are available at: \textbf{\artifactUrl{}}.


\balance
\bibliographystyle{ACM-Reference-Format}
\bibliography{biblio}

\end{document}